\documentstyle[prb,aps,graphicx]{revtex}
\begin{document}
\draft{}
\twocolumn[\hsize\textwidth\columnwidth\hsize
           \csname @twocolumnfalse\endcsname
%
\title{Torque magnetometry on single-crystal high temperature
superconductors near the critical temperature: a scaling
approach}
\author{J.~Hofer,
T.~Schneider,
J.M.~Singer,
M.~Willemin,
and H.~Keller}
\address{
Physik-Institut der Universit\"at Z\"urich, Winterthurerstr. 190,
CH-8057 Z\"urich, Switzerland \\
}
\author{T.~Sasagawa and K.~Kishio}
\address{
Department of Superconductivity, University of Tokyo, 7-3-1 Hongo,
Bunkyo-ku, Tokyo 113-8656, Japan \\
}
\author{K.~Conder and J.~Karpinski}
\address{
Laboratorium f\"ur Festk\"orperphysik, ETH H\"{o}nggerberg Z\"urich,
CH-8093 Z\"urich, Switzerland \\
}

\date{submitted to Phys. Rev. B, November 30 1999}
\maketitle
\begin{abstract}
Angular-dependent magnetic torque measurements performed near the critical
temperature on single crystals of
HgBa$_{2}$CuO$_{4+\mbox{\footnotesize{y}}}$,
La$_{2-\mbox{\footnotesize{x}}}$Sr$_{\mbox{\footnotesize{x}}}$CuO$_{4}$,
and YBa$_{2}$Cu$_{3}$O$_{6.93}$ are scaled, following the 3D $XY$
model, in order to determine the
scaling function $dG^{\pm}(z)/dz$ which describes the universal
critical properties near $T_{c}$.
A systematic shift of the
scaling function with increasing effective mass anisotropy $\gamma =
\sqrt{m_{ab}^{\ast}/m_{c}^{\ast}}$ is observed, which may be understood
in terms of a 3D-2D crossover. Further evidence for a 3D-2D crossover is found
from  temperature-dependent torque measurements
carried out in different magnetic fields at different field
orientations $\delta$, which show a quasi 2D ``crossing
region'' ($M^{\ast},T^{\ast}$). The occurrence of this
``crossing phenomenon'' is explained in a phenomenological way from the
weak $z$ dependence of the scaling function around a value $z = z^{\ast}$.
The ``crossing'' temperature $T^{\ast}$ is found to be
angular-dependent. Torque measurements above $T_{c}$ reveal that
fluctuations are strongly enhanced in the underdoped regime where the
anisotropy is large, whereas they are less important in the overdoped
regime.
\end{abstract}
%
]
%
\section{Introduction}
\label{section-introduction}

Angular-dependent magnetic torque measurements
per\-formed on a
HgBa$_{2}$CuO$_{4+\mbox{\footnotesize{y}}}$ single crystal around $T_{c}$
were recently described within a 3D $XY$ critical fluctuation model.
\cite{SchneiderEPJ98,HoferPRB99}
When 3D $XY$ critical fluctuations dominate, in a magnetic field
$\vec{B_{a}}$ the free energy
density is given by \cite{SchneiderEPJ98}
  \begin{equation}
  f = \frac{Q_{1}^{\pm}k_{B}T}{\xi_{a,0}^{\pm}\xi_{b,0}^{\pm}\xi_{c,0}^{\pm}
              |t|^{-3}}G^{\pm}(z).
  \label{f}
  \end{equation}
$\xi_{i,0}^{\pm}$ is the critical
amplitude of the correlation length along the $i$ direction, diverging as
$\xi_{i}^{\pm}(T) = \xi_{i,0}^{\pm}|t|^{-2/3}$ ($i = a,b,c$),
$t = T/T_{c} - 1$ is the reduced temperature, and $\pm = \mbox{sgn}(t)$.
$Q_{1}^{\pm}$
is an universal constant and $G^{\pm}(z)$ is an universal scaling
function.
It depends on the dimensionless variable $z$ which, if $\vec{B_{a}}$ is
applied
in the $ac$ plane of the sample, is given by
  \begin{equation}
   z = \frac{[\xi_{a}^{\pm}(T)]^{2} \gamma_{ab}
B_{a}}{\Phi_{0}}\epsilon(\delta).
   \label{z(delta)}
  \end{equation}
$\delta$ is the angle between $\vec{B_{a}}$ and the
$c$ axis of the
sample. $\epsilon(\delta) = (1/\gamma^{2} \sin^2\delta +
\cos^2 \delta)^{1/2}$, $\gamma =
\sqrt{m_{c}^{\ast}/m_{a}^{\ast}}$ is the effective mass anisotropy,
and $\gamma_{ab} =
\sqrt{m_{a}^{\ast}/m_{b}^{\ast}}$.

In a magnetic field applied in the $ac$ plane of a sample with volume $V$
the magnetic torque along the $b$ axis is given
by the derivative of $f$ with respect to the angle \cite{HoferPRB99}
  \begin{equation}
   \tau =
   \frac{VQ_{1}^{\pm}k_{B}TB_{a}\gamma }{2\Phi_{0}\xi_{a,0}^{\pm}
              |t|^{-2/3}}
              \left(1-\frac{1}{\gamma^{2}} \right)
              \frac{\sin(2\delta)}{\epsilon(\delta)}
              \frac{dG^{\pm}(z)}{dz}.
   \label{tau_scaling}
  \end{equation}
The functional form of the scaling function $dG^{\pm}/dz$
can be deduced in the following three limits:\cite{SchneiderEPJ98}
$dG^{-}/dz = C_{2,0}^{-} \ln(z)$ for $z \to 0^{-}$ ($T < T_{c}$, $B_{a} \to
0$),
$dG^{+}/dz = C_{0}^{+} z$ for $z \to 0^{+}$ ($T > T_{c}$, $B_{a} \to 0$),
$dG^{\pm}/dz = C_{\infty}^{\pm} \sqrt{z}$ for
$z \to \infty$ ($B_{a} \neq 0$, $T \to T_{c}$). In the intermediate $z$-regime
$dG^{\pm}/dz$ was determined from angular-dependent measurements
using a scaling procedure dictated by Eq.~(\ref{tau_scaling}), as
described in detail in Ref.~\onlinecite{HoferPRB99}.
In this paper we apply this
scaling approach to single crystals of
HgBa$_{2}$CuO$_{4+\mbox{\footnotesize{y}}}$,
La$_{2-\mbox{\footnotesize{x}}}$Sr$_{\mbox{\footnotesize{x}}}$CuO$_{4}$,
and YBa$_{2}$Cu$_{3}$O$_{6.93}$
to explore the limitations of the approach arising from a large
anisotropy.
We assume $\gamma_{ab} = 1$ for
HgBa$_{2}$CuO$_{4+\mbox{\footnotesize{y}}}$ and for
La$_{2-\mbox{\footnotesize{x}}}$Sr$_{\mbox{\footnotesize{x}}}$CuO$_{4}$,
which is reasonable due to the tetragonal structure of these
compounds. For YBa$_{2}$Cu$_{3}$O$_{6.93}$ we
use $\gamma_{ab} = 1.12$ (Ref.~\onlinecite{WilleminPRL98}).

Due to the large effective mass anisotropy most cuprates are
in a 3D-2D crossover regime at temperatures slightly below $T_{c}$
where $\xi_{c}(T) = \xi_{ab}(T) / \gamma$ becomes smaller than the
interlayer distance $s$.\cite{SchneiderPC99}
In the samples investigated the anisotropy values
reach up to $\gamma \approx 50$. These large
$\gamma$ values may lead to a shrinking of the temperature
regime where the 3D scaling approach is applicable. The
observation of a so-called ``crossing point'' in the temperature dependence
of the magnetization $M$ can test the
dimensionality of a high-$T_{c}$ material. For a 3D system the quantity
$M/B_{a}^{1/2}$ is predicted to be field-independent at the critical
temperature $T_{c}$. On the other hand, in a 2D system the
magnetization $M = M^{\ast}$ is predicted to be field-independent
at the critical temperature of this system $T^{\ast} =
T_{KT}$ (Kosterlitz-Thouless transition temperature).\cite{SchneiderPC99}
In fact, YBa$_{2}$Cu$_{3}$O$_{6.93}$ with a moderate anisotropy $\gamma
\approx 7$ only shows a 3D ``crossing point'' at $T = T_{c}$
(Ref.~\onlinecite{JunodPC94}), whereas
Bi$_{2.15}$Sr$_{1.85}$CaCu$_{2}$O$_{8+\mbox{\footnotesize{y}}}$ with
$\gamma \approx 150$ shows a rather well defined 2D ``crossing point''
($M^{\ast},T^{\ast}$) at $T = T^{\ast} < T_{c}$.\cite{JunodPC94,KesPRL91} In
the same compound 2D effects have been observed at low
temperatures in muon spin rotation measurements.\cite{AegerterPRB98}
The 2D ``crossing point'' ($M^{\ast},T^{\ast}$) was observed in
Bi$_{2.15}$Sr$_{1.85}$CaCu$_{2}$O$_{8+\mbox{\footnotesize{y}}}$
for fields applied along the $c$ axis,
but not for fields in the $ab$ plane.\cite{SchneiderPC99} We
performed temperature-dependent torque measurements on a
La$_{1.914}$Sr$_{0.086}$CuO$_{4}$
crystal with $\gamma = 46$ to investigate the 2D ``crossing point''
in this material.
The measurements were carried out at different angles $\delta$ in order
to clarify the angular dependence of the ``crossing'' phenomenon.

Although an enhanced anisotropy may lead to a shrinking of the
the 3D scaling temperature region, it increases the importance of
fluctuation effects.\cite{FisherPRB91,GoldenfeldBOOK92}
Thus, fluctuation effects are assumed to be more important in
underdoped samples showing a large anisotropy. From angular-dependent
torque measurements performed above $T_{c}$
we determined the doping dependence of fluctuation effects in
La$_{2-\mbox{\footnotesize{x}}}$Sr$_{\mbox{\footnotesize{x}}}$CuO$_{4}$
compounds with different Sr contents $x$ ranging from the underdoped
to the overdoped regime.

\section{Experimental details}
\label{section-experiment}

The HgBa$_{2}$CuO$_{4+\mbox{\footnotesize{y}}}$ single crystals
were grown using a high-pressure
growth technique.\cite{KarpinskiSST99} Oxygen annealing resulted in
two crystals with $y = 0.096$ and $y = 0.108$, respectively. The
La$_{2-\mbox{\footnotesize{x}}}$Sr$_{\mbox{\footnotesize{x}}}$CuO$_{4}$
samples were cut from single crystals grown by the
traveling-solvent-floating-zone method with different nominal Sr
compositions.\cite{KimuraPC92,SasagawaPRL98} The Sr contents $x$ of the
samples were
determined by an
electron-probe-micro-analysis to be $x = 0.080$, $0.086$, $0.132$,
$0.146$, and $0.180$, respectively.
Angular-dependent torque data obtained on the YBa$_{2}$Cu$_{3}$O$_{6.93}$
sample used in Ref.~\onlinecite{WilleminPRL98} were reanalyzed for
this work.

The HgBa$_{2}$CuO$_{4+\mbox{\footnotesize{y}}}$ and
La$_{2-\mbox{\footnotesize{x}}}$Sr$_{\mbox{\footnotesize{x}}}$CuO$_{4}$
single crystals had typical dimensions of $150 \times 150 \times 50
\mbox{ } \mu \mbox{m}^{3}$ ($m \approx 8 \mbox{ } \mu$g). They were mounted
on a
piezoresistive torque sensor between the poles of a conventional
NMR-magnet with a maximum field of 1.5 T.\cite{WilleminJAP98} The
magnetic torque $\vec{\tau} = V \vec{M} \times \vec{B_{a}}$ acting on
the magnetization $\vec{M}$ of the sample was measured as a function of the
angle $\delta$ at different constant field strengths and
temperatures below and above $T_{c}$. Only
data showing reversible behavior were used for the analysis. The reversible
regime was partly extended using an AC field
perpendicular to $\vec{B_{a}}$ to enhance relaxation processes.
\cite{WilleminPRB98}
From these reversible angular-dependent torque data
the scaling function $dG^{\pm}/dz$ was determined for each sample.

Temperature-dependent measurements were per\-for\-med on the
La$_{1.914}$Sr$_{0.086}$CuO$_{4}$ crystal for fields $B_{a} = 0.25$,
$0.50$, $0.75$, $1.00$, $1.20$, and $1.40$ T at angles $\delta = 15^{\circ}$,
$30^{\circ}$, $45^{\circ}$, $60^{\circ}$, $75^{\circ}$, and $85^{\circ}$
to study the ``crossing point'' phenomenon. The torque signal
was continuously recorded upon cooling the sample in the applied field at a
cooling rate of $dT/dt \simeq -0.05$ K/s. The measurements
were only possible by compensating the strongly temperature-dependent
background of the piezoresistive torque sensor by a second cantilever
placed near the sample.\cite{WilleminJAP98} In fact, in the
temperature regime $18.8 \mbox{ K} < T < 25.2 \mbox{ K}$, where these
measurements were performed, no temperature-dependent signal was
observed in zero field, pointing towards an almost perfect
compensation.

The same compensation method was also used for the
angular-dependent measurements where it reduced the angular-dependent
magnetoresistive background of the cantilever drastically. The
La$_{2-\mbox{\footnotesize{x}}}$Sr$_{\mbox{\footnotesize{x}}}$CuO$_{4}$
samples with concentrations $x = 0.080,\ 0.146, \mbox{ and } 0.180$
were measured using the cantilever in the so-called torsion mode where
a torsion of the sensor around its long axis is measured by a
difference in the resistance of the two piezoresistive paths
integrated in the outer two legs of the lever.\cite{WilleminJAP98}
In this differential mode no additional compensation lever is needed,
and no angular dependence
of the magnetoresistance is observed due to the fact that the
magnetic field is always perpendicular to the main current path of the
cantilever. This was especially advantageous for angular-dependent
measurements, performed above $T_{c}$ in order to determine the
temperature $T_{0}$ where the diamagnetic
signal arising from superconducting fluctuations exactly compensates the
normal paramagnetic background of the sample.

\section{Scaling of the angular-dependent data}

\label{section-scaling}
\begin{figure}[t]
        \centering
        \includegraphics[width=0.8\linewidth]{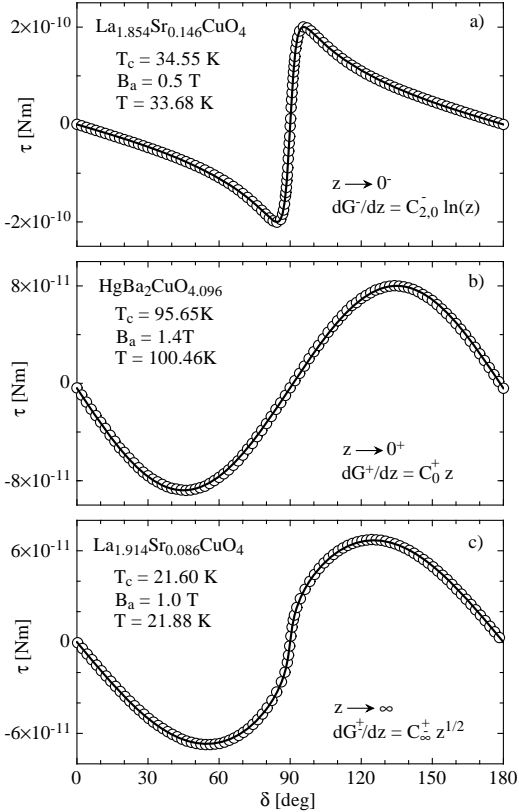}
        \caption[~]{Angular-dependent torque data recorded in the
        three limits where the analytical form of $dG^{\pm}/dz$ is
        known. The data are
        well described by the limiting behavior of the scaling
        function (solid lines). For clarity not all measured data
        points are shown. a) measurement on
        La$_{1.854}$Sr$_{0.146}$CuO$_{4}$ in $B_{a} = 0.5$ T, $T =
        33.68$ K ($z \to 0^{-}$). b) measurement on
        HgBa$_{2}$CuO$_{4.096}$ in $B_{a} = 1.4$ T, $T =
        100.46$ K ($z \to 0^{+}$).  c) measurement on
        La$_{1.914}$Sr$_{0.086}$CuO$_{4}$ in $B_{a} = 1.0$ T, $T =
        21.88$ K ($z \to \infty)$.}
	\protect\label{limits}
\end{figure}

\begin{figure}[tb]
        \centering
        \includegraphics[width=0.8\linewidth]{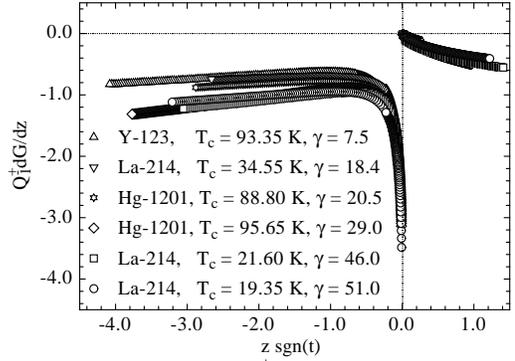}
        \caption[~]{Scaling function $dG^{\pm}/dz$ determined by
        scaling angular-dependent data for
        YBa$_{2}$Cu$_{3}$O$_{6.93}$, La$_{1.854}$Sr$_{0.146}$CuO$_{4}$,
        HgBa$_{2}$CuO$_{4.108}$, HgBa$_{2}$CuO$_{4.096}$,
        La$_{1.914}$Sr$_{0.086}$CuO$_{4}$, and
        La$_{1.920}$Sr$_{0.080}$CuO$_{4}$. The qualitative behavior of
        $dG^{\pm}/dz$ is the same for all samples. However, below
        $T_{c}$ a systematic
        shift with increasing $\gamma$ is observed.}
	\protect\label{dG/dz(all)}
\end{figure}

Angular-dependent torque data recorded in the three limits $z \to
0^{-}$, $z \to 0^{+}$, and $z \to \infty$, where the
analytical form of $dG^{\pm}/dz$ can be deduced, are displayed in
Fig.~\ref{limits}.
The data are well described by the limiting behavior of the
scaling function in the corresponding $z$ limits.
For six samples with varying anisotropy $\gamma$ the scaling procedure
outlined in
Ref.~\onlinecite{HoferPRB99} was applied to the angular-dependent
data in order to determine the scaling function $dG^{\pm}/dz$.
The normal paramagnetic background of the sample was determined from
measurements performed above $T_{c}$ (cf. Section~\ref{section-T>Tc})
and subtracted from the data. The effective mass anisotropy $\gamma$ and the
critical amplitude of the correlation length $\xi_{a,0}^{-}$ were
determined from angular-dependent measurements performed below $T_{c}$
in the low $z$ regime where $dG^{-}/dz = C_{2,0}^{-} \ln(z)$
[Fig.~\ref{limits}(a)]. In this
limit the 3D $XY$ form of the angular dependence of the torque coincides
with the prediction of the 3D anisotropic London model.\cite {FarrellPRL88}
Equation~(\ref{tau_scaling}) was then resolved for $Q_{1}^{\pm}
dG^{\pm}/dz$ using the values obtained for $\gamma$ and
$\xi_{a,0}^{-}$. $T_{c}$ was iteratively adjusted in steps of $0.05$ K in
order to get
the best overlap of scaled $dG^{-}/dz$ curves.
For $T > T_{c}$, $\xi_{a,0}^{+}$ was calculated using the universal relation
$\xi_{a,0}^{+} = (R^{+}/R^{-})(A^{-}/A^{+})^{1/3}
\xi_{a,0}^{-} \simeq 0.38 \times \xi_{a,0}^{-}$.
\noindent
The specific heat universal
constants $R^{+} = 0.36$, $R^{-} = 0.95$
and critical amplitudes ratio $A^{+}/A^{-} \simeq 1$
were taken from Ref.~\onlinecite{BervillierPRB76}. The
scaling functions obtained for all six samples are plotted in
Fig.~\ref{dG/dz(all)} as a function of $z \cdot \mbox{sgn}(t)$.

The qualitative behavior of $dG^{\pm}/dz$ is the same for all samples.
Especially, for $T < T_{c}$ the crossover from the
$\ln(z)$ dependence to the $\sqrt{z}$ dependence is clearly seen for
all compounds. It occurs around $z \cdot \mbox{sgn}(t) \simeq -1$.
All scaled data lie within an interval of 80\%. This
scattering partly reflects the experimental errors in
performing the scaling procedure. An initial uncertainty arises from the
paramagnetic background of the sample which is determined at
temperatures far above $T_{c}$ and assumed to adopt the same value at
temperatures around $T_{c}$. For temperatures very close to $T_{c}$, where
the torque signal arising from superconductivity is of the same order
of magnitude as the paramagnetic background,
any uncertainty in determining this background can have a rather large
effect. Secondly, inhomogeneities within the samples will lead to
finite size effects. Talking in terms of a distribution of $T_{c}$,
$|dG^{\pm}/dz|$ will be overestimated if
\noindent
the scaling
procedure is performed using one single $T_{c}$ value.
This overestimation comes from parts of the sample having a slightly higher
$T_{c}$ and therefore giving a larger contribution to the measured torque
signal.
Indeed, as seen in Fig.~\ref{dG/dz(all)} $|dG^{\pm}/dz|$ obtained for
YBa$_{2}$Cu$_{3}$O$_{6.93}$, the sample with the sharpest transition,
adopts the smallest values. A third uncertainty is coming from
the volume $V$ which is somewhat problematic to determine for such
small samples. Finally, a crucial parameter for the scaling procedure is the
correlation length amplitude $\xi_{a,0}^{-}$ which has to be
determined from measurements performed in the limit $z \to 0^{-}$.
Since torque measurements cannot be carried out in very low magnetic
fields where the sensitivity is not high enough, the only possibility
to reach this limit is to reduce the temperature.
The width of the 3D $XY$ scaling temperature regime, where the temperature
dependence of the correlation length is given as $\xi_{a}^{-}(T) =
\xi_{a,0}^{-}|t|^{-2/3}$, is determined by corrections to scaling.
For anisotropic systems, being in a 3D-2D crossover regime, the
amplitudes of the correction terms can become rather large.\cite{FisherRMP74}
This may lead to a shrinking of the temperature regime where 3D
scaling is applicable.
The correction amplitudes are not
known a priori and vary between
different samples,
\begin{table}
    \caption[~]{Critical amplitudes determined by fitting the
    angular-dependent data recorded in the limit $z \to 0^{-}$.
    For La$_{1.854}$Sr$_{0.146}$CuO$_{4}$,
La$_{1.914}$Sr$_{0.086}$CuO$_{4}$,
    and La$_{1.920}$Sr$_{0.080}$CuO$_{4}$ the tendency for $\gamma$,
    $\xi_{a,0}^{-}$, and
    $\lambda_{a,0}$ to increase with decreasing doping in the underdoped
    regime is evident.}
\begin{tabular}{lrrrr}
	sample & $T_{c}$ [K] & $\gamma$ & $\xi_{a,0}^{-}$ [\AA]&
$\lambda_{a,0}$ [\AA]\\
	\hline
	YBa$_{2}$Cu$_{3}$O$_{6.93}$       & 93.35(5) & 7.5(5)  & 13(2)   &
	1130(90) \\
	HgBa$_{2}$CuO$_{4.108}$           & 88.80(5) & 20.5(8) & 29(4)   &
	1000(80) \\
	HgBa$_{2}$CuO$_{4.096}$           & 95.65(5) & 29(1)   & 26(4)   &
	780(80) \\
    La$_{1.854}$Sr$_{0.146}$CuO$_{4}$ & 34.55(5) & 18.4(6) & 30(4)   &
    1720(130) \\
    La$_{1.914}$Sr$_{0.086}$CuO$_{4}$ & 21.60(5) & 46(1)   & 79(12)  &
    2230(170) \\
    La$_{1.920}$Sr$_{0.080}$CuO$_{4}$ & 19.35(5) & 51(1)   & 101(15) &
    2530(190) \\
\end{tabular}
    \protect\label{ampl}
\end{table}
\begin{figure}[t]
        \centering
        \includegraphics[width=0.8\linewidth]{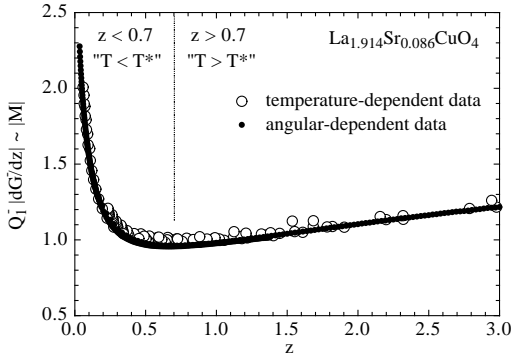}
        \caption[~]{Scaling function $|dG^{-}/dz|$ for temperatures $T <
        T_{c}$ for the case of La$_{1.914}$Sr$_{0.086}$CuO$_{4}$,
        derived from temperature-dependent (large symbols) and from
        angular-dependent data (small symbols).
        The magnetization $M$ shows a crossover in its $z$ dependence
        around $z = 0.7$. A ``crossing region''
        ($M^{\ast},T^{\ast}$) is expected to be located around
        $z^{\ast} \simeq 0.7$.}
	\protect\label{dG/dz(La086)}
\end{figure}
\noindent
therefore only the
3D $XY$ temperature dependence
was taken into account for our analysis. The critical amplitude of the
correlation length determined from measurements performed at different
temperatures and fields, such that $z \to 0^{-}$ was rather well fulfilled for
each measurement, showed a scattering of about 20\%. Therefore, we
estimate the error in
$\xi_{a,0}^{-}$ to be 15\%, leading to the same error in $dG^{\pm}/dz$
and to an error of 30\% in $z$.

As observed in Fig.~\ref{dG/dz(all)}, the scaling functions
obtained for the samples investigated deviate systematically with
increasing $\gamma$ from that of YBa$_{2}$Cu$_{3}$O$_{6.93}$, the sample
with the smallest anisotropy. The amplitudes
of the correction terms
mentioned above depend on how gradually the crossover occurs
\cite{FisherRMP74}, and $\gamma$ is a measure for quasi 2D behavior of
the cuprates. Therefore, it is plausible that the corrections to 3D scaling
depend systematically on the anisotropy are getting more important with
increasing $\gamma$.
However, there are not enough data available to determine these
corrections quantitatively.

The critical amplitudes obtained for all six samples from the 3D $XY$ scaling
approach are listed in Table~\ref{ampl}. The penetration depth
amplitude $\lambda_{a,0}$ is calculated using the universal relation
$k_{B}T_{c} = \Phi_{0}^{2}
\xi_{a,0}^{-}/(4\pi^{2}\mu_{0}\lambda_{a,0}^{2}\gamma)$.\cite{SchneiderEPJ98}
An error of 15\% was assumed for $\xi_{a,0}^{-}$ (see above) leading
to an error of $\sim 8$\% for $\lambda_{a,0}$.
The tendency for $\gamma$, $\xi_{a,0}^{-}$, and
$\lambda_{a,0}$ to increase with decreasing doping in the underdoped
regime, as observed in different experiments,
\cite{ChienPC94,HoferPC98,JaccardEPL96,SchneiderJMP93} is evident
for the
La$_{2-\mbox{\footnotesize{x}}}$Sr$_{\mbox{\footnotesize{x}}}$CuO$_{4}$
samples.
The anisotropy values found for
La$_{2-\mbox{\footnotesize{x}}}$Sr$_{\mbox{\footnotesize{x}}}$CuO$_{4}$
are in rather good
\begin{table}
    \caption[~]{Universal constants for the three limits $z \to
    0^{-}$, $z \to 0^{+}$, and $z \to \infty$ determined by scaling
    the angular-dependent data according to Eq.~(\ref{tau_scaling}).}
\begin{tabular}{ll}
	universal constant & range  \\
	\hline
	$Q_{1}^{-}C_{2,0}^{-}$    & 0.6 to 0.8  \\
	$Q_{1}^{+}C_{0}^{+}$      & -0.8 to -1.0 \\
	$Q_{1}^{+}C_{\infty}^{+}$ & -0.3 to -0.5 \\
\end{tabular}
    \protect\label{const}
\end{table}
\noindent
agreement with normal state
anisotropies $\gamma_{\rho} = \sqrt{\rho_{c}/\rho_{ab}}$ in compounds
with similar doping.\cite{KimuraPC92,SasagawaPRL98}

Table~\ref{const} lists the range which we determined from the 3D $XY$
scaling approach for the three universal constants
corresponding to the three limits of $z$ (cf. Fig.~\ref{limits}).
The largest uncertainty is
found for the limit $z \to \infty$, i.e. $T \to T_{c}$. This may
reflect the problems arising from a distribution of $T_{c}$, as already
discussed above.

\section{The ``crossing point'' phenomenon}
\label{section-crossing}

Taking the derivative of $f$ [Eq.~(\ref{f})] with respect to the magnetic
field, we find the 3D-$XY$ expression for the magnetization
  \begin{equation}
  M = \frac{Q_{1}^{\pm}k_{B}T\gamma}{\Phi_{0}\xi_{a,0}^{\pm}
              |t|^{-2/3}}
              \frac{\left(1/\gamma^{4} \sin^2\delta +
              \cos^2 \delta \right)^{1/2}}
              {\epsilon(\delta)}
              \frac{dG^{\pm}(z)}{dz}.
  \label{M}
  \end{equation}
In the limit $z \to \infty$, where $dG^{\pm}/dz \propto \sqrt{z}
\propto \sqrt{B_{a}}$, Eq.~(\ref{M}) predicts the 3D ``crossing point''
($M/B_{a}^{1/2},T_{c}$) which is observed in the temperature-dependent
magnetization.\cite{JunodPC94} Here we concentrate on the 2D ``crossing
point''
($M^{\ast},T^{\ast}$), which has been widely
studied.\cite{SchneiderPC99,JunodPC94,KesPRL91} For $T < T^{\ast}$,
$|M|$ decreases with increasing field, whereas for $T > T^{\ast}$ it
increases with increasing field. The same change in
the field dependence of the magnetization is described by the
scaling function $dG^{-}/dz$ for $T < T_{c}$.
$|dG^{-}/dz| \propto |M|$,
as determined by scaling the angular-dependent data, is plotted in
Fig.~\ref{dG/dz(La086)} for the case of
La$_{1.914}$Sr$_{0.086}$CuO$_{4}$ (small symbols). The scaling function
obtained from temperature-dependent data measured at different angles and
fields is also included (large symbols). The agreement of the
angular-dependent and temperature-dependent measurements is excellent.
For $z < 0.7$, $|dG^{-}/dz|$ decreases with increasing $z$, whereas for
$z > 0.7$ it increases with increasing $z$. Let us recall the
dependence of $z$ on temperature, field and angle by looking at
Eq.~(\ref{z(delta)}).
$z$ can be increased in three ways:
(i) $T \to T_{c}$, $B_{a} = const$, $\delta = const$;
(ii) $B_{a} \to \infty$, $T = const$, $\delta = const$;
(iii) $\delta \to 0^{\circ}$, $T = const$, $B_{a} = const$. Since $z \propto
B_{a}$,
we recover the above crossover in the field dependence of $M$ from
$dG^{-}/dz$,
if for an intermediate field of $B_{a} = 1$ T the scaling argument $z$
adopts the value $0.7$, where $|dG^{-}/dz|$ has its minimum,
at the temperature $T = T^{\ast}$. The situation
is indicated by the vertical line in Fig.~\ref{dG/dz(La086)}. Thus, we can
define in a purely phenomenological way a ``crossing temperature''
$T^{\ast}$ by the condition
  \begin{equation}
  z^{\ast} = z(T^{\ast}, B_{a} = 1 \mbox{ T}) = 0.7.
  \label{z*}
  \end{equation}
The ``crossing'' magnetization is then given by $M^{\ast} =
M(T^{\ast},B_{a} = 1\mbox{ T})$. From Fig.~\ref{dG/dz(La086)} it is
evident that we will not observe a perfect ``crossing point''
($M^{\ast},T^{\ast}$), where $M$ is exactly independent of the applied
field. Indeed, at the temperature $T^{\ast}$ as defined by
Eq.~(\ref{z*}), for all applied fields $B_{a} > 1$ T, $z > z^{\ast}$
and therefore $|M(T^{\ast})| > |M^{\ast}|$.
\begin{figure}[htb]
        \centering
        \includegraphics[width=0.8\linewidth]{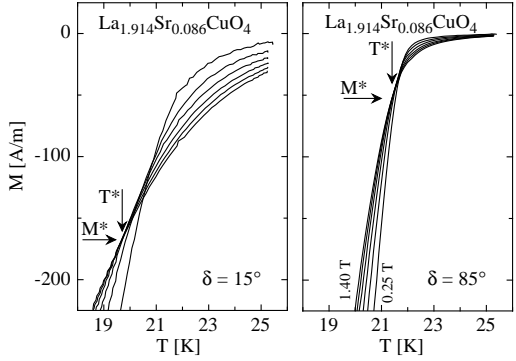}
        \caption[~]{Temperature-dependent measurements performed on
        La$_{1.914}$Sr$_{0.086}$CuO$_{4}$ in
        different fields $B_{a} = 0.25$, 0.50, 0.75, 1.00,
        1.20, and 1.40 T (from bottom to top for
        $T < T^{\ast}$) for field orientations
        $\delta = 15^{\circ}$ (left panel)
        and $\delta = 85^{\circ}$ (right panel), respectively.
        $M^{\ast}$ as well
        as $T^{\ast}$ depend on $\delta$.
        The ``crossing point'' is better defined for a field orientation
        close to the $ab$ plane.}
	\protect\label{crossing}
\end{figure}
 \begin{figure}[htb]
        \centering
        \includegraphics[width=0.8\linewidth]{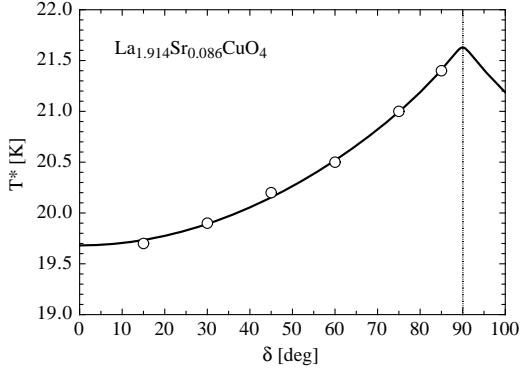}
        \caption[~]{Measured ``crossing temperature'' $T^{\ast}$ as a function
        of the angle $\delta$ for La$_{1.914}$Sr$_{0.086}$CuO$_{4}$ (open
        circles).
        $T^{\ast}$ is increasing when the field is approaching
        the $ab$ plane.
        The solid line represents a fit to Eq.~(\ref{T*}) with
        parameters $T_{c} = 21.75(5)$ K, $T^{\ast}(0^{\circ}) =
        19.68(3)$ K.}
	\protect\label{T*(delta)}
\end{figure}
 \begin{figure}[htb]
        \centering
        \includegraphics[width=0.8\linewidth]{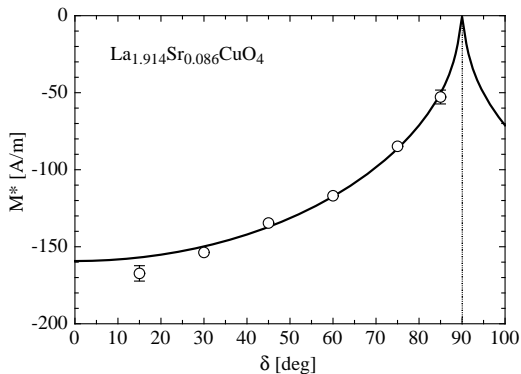}
        \caption[~]{Measured ``crossing magnetization'' $M^{\ast}$ as a
function
        of the angle $\delta$ for La$_{1.914}$Sr$_{0.086}$CuO$_{4}$ (open
        circles).
        The solid line is calculated without any free parameter, using
        Eqs.~(\ref{M}) and (\ref{T*}).}
	\protect\label{M*(delta)}
\end{figure}
\noindent
Since $|M|$ decreases
monotonically upon approaching $T_{c}$ from below, only for a
temperature $T' > T^{\ast}$ the magnetization $M$ adopts the value
$M^{\ast}$. However, $|dG^{-}/dz|$ only shows a weak
$z$ dependence for $z > z^{\ast}$ and $T'$ is close to $T^{\ast}$.
Therefore, if one looks at temperature-dependent magnetization measurements
over a rather large temperature interval, a quite well defined
``crossing point'' will be observed for fields up to $5$ T.
The situation is slightly different for fields $B_{a} < 1$ T. In that
case $z(T^{\ast}) < z^{\ast}$, and again $|M(T^{\ast})| > |M^{\ast}|$.
But $|dG^{-}/dz|$ shows a strong $z$ dependence for $z < 0.2$.
Therefore $T'$, where $M(T') = M^{\ast}$, can be considerably
higher than $T^{\ast}$, and magnetization curves measured in
fields $B_{a} < 0.3$ T tend to ``cross'' at higher temperatures. Thus we
expect to observe a ``crossing region'' with a width that is mainly
determined by low field measurements.
    In fact, experiments show that the ``crossing point'' is actually a
``crossing region'', and low field measurements lie at higher
temperatures.\cite{SchneiderPC99}

If we take the angular dependence of $z$ into account,
we see that for a field orientation close to the $ab$ plane of the
sample, where $z(\delta)$ has a minimum, Eq.~(\ref{z*}) is fulfilled for
a higher temperature $T^{\ast}$. Therefore, we expect
$T^{\ast}$ to increase with increasing $\delta$ in the range
$0^{\circ} < \delta < 90^{\circ}$. Furthermore, as the temperature
dependence of $|M(z < z^{\ast})|$ gets stronger upon approaching $T_{c}$
the width of the crossing region, which is determined by
measurements with small $z$, is expected to shrink for field orientations
close to the $ab$ plane. This trend is clearly seen in Fig.~\ref{crossing}
where temperature-dependent measurements on
La$_{1.914}$Sr$_{0.086}$CuO$_{4}$ performed in different fields are
presented for $\delta = 15^{\circ}$ and $\delta = 85^{\circ}$,
respectively. As indicated by the arrows, $M^{\ast}$ and $T^{\ast}$
are determined from the crossing of the magnetization curves recorded
in fields $0.75 \mbox{ T} < B_{a} < 1.40$ T. The measurements at
$B_{a} = 0.25$ T and $0.50$ T are not taken into account.

Presuming $z^{\ast}(\delta) = z^{\ast}(0^{\circ})$
for all orientations $\delta$, one obtains
from Eq.~(\ref{z(delta)}) an angular-dependent ``crossing temperature''
  \begin{equation}
  T^{\ast}(\delta) = [\epsilon(\delta)]^{3/4}T^{\ast}(0^{\circ}) +
                     T_{c}(1 - [\epsilon(\delta)]^{3/4}).
  \label{T*}
  \end{equation}
Figure~\ref{T*(delta)} shows $T^{\ast}$ as a function of
the angle $\delta$. The solid line represents a least square fit
to Eq.~(\ref{T*}).
The observed angular dependence of $T^{\ast}$ is well reproduced
with the fitting parameters $T_{c} = 21.75(5)$ K and
$T^{\ast}(0^{\circ}) = 19.68(3)$ K (the anisotropy was fixed at
$\gamma = 46$). With $T^{\ast}(0^{\circ}) = 19.68$ K we
find the actual value of $z^{\ast}$ to be
$z(T^{\ast}(0^{\circ}),\delta = 0^{\circ}, B_{a} = 1 \mbox{ T}) = 0.69$.
Indeed, the ``crossing region'' is located around the $z$-value,
where $|dG^{-}/dz|$ adopts its minimum. Comparing the left panel to the
right panel of Fig.~\ref{crossing}, we can see that $|M^{\ast}|$ adopts
a considerable higher value for $\delta = 15^{\circ}$ than for
$\delta = 85^{\circ}$.
Figure~\ref{M*(delta)} displays $M^{\ast}$ as a function of $\delta$.
The solid line is
calculated by combining Eqs.~(\ref{M}) and (\ref{T*}) using
$T_{c} = 21.75$ K, $T^{\ast}(0^{\circ}) = 19.68$ K, $\xi_{a,0}^{-} = 79\ \AA$,
$\gamma = 46$, and
$Q_{1}^{-}dG^{-}/dz = -1$ for $z = z^{\ast}$ (see Fig.~\ref{dG/dz(La086)}).
The calculated angular-dependence of $M^{\ast}$ is in fair agreement
\begin{figure}[htb]
        \centering
        \includegraphics[width=0.8\linewidth]{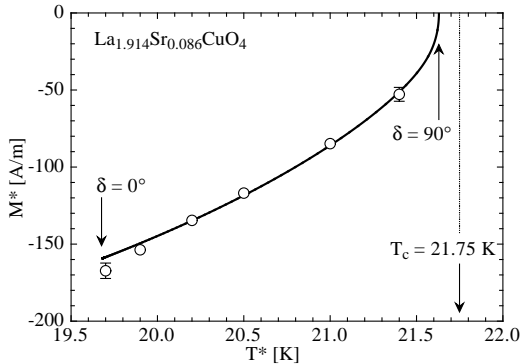}
        \caption[~]{Measured ``crossing magnetization'' $M^{\ast}$ as a
function
        of $T^{\ast}$ for La$_{1.914}$Sr$_{0.086}$CuO$_{4}$(open
        circles).
        The solid line is combined from the curves shown in
        Figs.~\ref{T*(delta)} and \ref{M*(delta)}. Note the extremely small
        value $M^{\ast}(90^{\circ}) = -0.6$ A/m for fields
        applied in the $ab$ plane.}
	\protect\label{M*(T*)}
\end{figure}
\noindent
with the measured data.
Figure~\ref{M*(T*)} displays the dependence of $M^{\ast}$ on $T^{\ast}$
in the La$_{1.914}$Sr$_{0.086}$CuO$_{4}$ sample. Upon turning the
field towards the $ab$ plane, $T^{\ast}$ monotonically increases from
$19.68$ K to $21.63$ K (see also Fig.~\ref{T*(delta)}). Due to the
strong temperature dependence of the magnetization in this
temperature range, $M^{\ast}$ changes dramatically. For
$\delta = 0^{\circ}$ we find $M^{\ast} \simeq -165$ Am$^{2}$, whereas it
almost vanishes for $\delta = 90^{\circ}$,
where we calculate $M^{\ast}(\delta = 90^{\circ}) =
-0.6$ A/m, $T^{\ast}(\delta = 90^{\circ}) = 21.63$ K. Due to the small value
of $M^{\ast}$, it is very difficult to
resolve the ``crossing point'' for fields applied at $\delta =
90^{\circ}$. Indeed, magnetization curves recorded at $\delta =
90^{\circ}$ are found to join at $T_{c}$ rather
than to ``cross'' slightly below $T_{c}$
(Ref.~\onlinecite{SchneiderPC99}).

\section{Torque signal above the critical temperature:
         importance of superconducting fluctuations}
\label{section-T>Tc}

\begin{figure}[htb]
        \centering
        \includegraphics[width=0.8\linewidth]{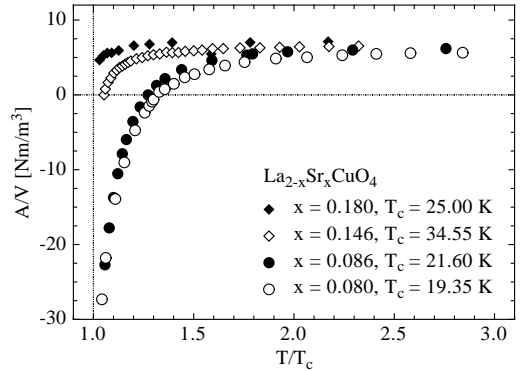}
        \caption[~]{Normalized $\sin(2\delta)$ amplitude $A$ as a function of
        reduced temperature $T/T_{c}$ for

La$_{2-\mbox{\footnotesize{x}}}$Sr$_{\mbox{\footnotesize{x}}}$CuO$_{4}$
        single crystals with different doping $x$. For the underdoped samples
        ($x = 0.080$ and $0.086$)
        fluctuation effects can be observed already at temperatures
        much higher than $T_{c}$}
	\protect\label{ampl(T)}
\end{figure}

The temperature dependence of the amplitude $A$ of the $\sin(2\delta)$ signal
observed for temperatures $T > T_{c}$ [Fig.~\ref{limits}(b)] yields
information about the
importance of superconducting fluctuations. Figure~\ref{ampl(T)} shows
the normalized amplitude $A/V$ as a function of the reduced
temperature $T/T_{c}$ for
La$_{2-\mbox{\footnotesize{x}}}$Sr$_{\mbox{\footnotesize{x}}}$CuO$_{4}$
samples with different doping $x$. At temperatures well
above $T_{c}$ a positive, almost temperature-independent amplitude
$A$ is observed. This signal is mainly due to the Van Vleck
paramagnetic response of the copper ions.\cite{LeePRB90,MehranPRB90}
Within the investigated doping range
the Van Vleck contribution is found to be independent of $x$.
In the underdoped regime, below temperatures $T \approx 2 \cdot T_{c}$,
$A$ starts to decrease
and finally vanishes at a temperature $T_{0} > T_{c}$.
For $T < T_{0}$ the
$\sin(2\delta)$ amplitude is negative and $|A|$ increases
dramatically upon approaching $T_{c}$. The observed temperature dependence
of $A$ is due to superconducting fluctuations. This is evidenced by
the fact that the signal is clearly related with $T_{c}$. Moreover,
the temperature dependence of $A$ is well described by the 3D $XY$
expression for
\noindent
temperatures close to $T_{c}$
(Ref.~\onlinecite{HoferPRB99}). With
$dG^{+}/dz = C_{0}^{+} z$, from Eq.~(\ref{tau_scaling}) we find
  \begin{equation}
  \frac{A(T)}{V} = \frac{Q_{1}^{+}C_{0}^{+}k_{B}T_{c}B_{a}^{2}}{2\Phi_{0}^{2}}
                \xi_{a,0}^{+}\gamma \left(1 - \frac{1}{\gamma^{2}} \right)
                \frac{T}{T_{c}} \left(\frac{T}{T_{c}} - 1 \right)^{-2/3}.
  \label{A/V}
  \end{equation}
It is evident from this expression that $A/V$ is large (i.e.
fluctuations are
important) for large values of $\gamma$ and $\xi_{a,0}^{+}$. As seen from
Table~\ref{ampl}, $\gamma$ and $\xi_{a,0}^{+}$ both increase with decreasing
doping $x$ in underdoped
La$_{2-\mbox{\footnotesize{x}}}$Sr$_{\mbox{\footnotesize{x}}}$CuO$_{4}$.
The decrease in $T_{c}$, which also affects $A/V$, is smaller than the
increase in $\gamma$ and $\xi_{a,0}^{+}$. Thus, from Eq.~(\ref{A/V})
we can conclude that fluctuations are important in the underdoped
regime. More general, the importance of fluctuations in highly
anisotropic systems is seen from the fact, that an anisotropic system
can be rescaled to the isotropic case with a strongly enhanced
temperature $T' = \gamma T$.\cite{BlatterRMP94} This enhancement of the
rescaled temperature gives rise to strong fluctuations.

\begin{figure}[htb]
        \centering
        \includegraphics[width=0.8\linewidth]{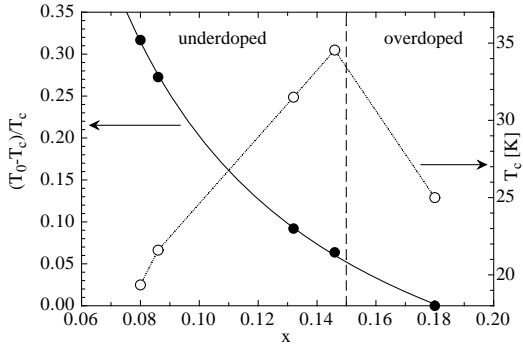}
        \caption[~]{Normalized difference $(T_{0} - T_{c})/T_{c}$ and
        $T_{c}$ as a
        function of doping $x$ for

La$_{2-\mbox{\footnotesize{x}}}$Sr$_{\mbox{\footnotesize{x}}}$CuO$_{4}$.
        $T_{0}$ is the temperature, where the fluctuation signal
        completely compensates the paramagnetic background of the sample.
        The importance of superconducting fluctuations clearly
        increases for decreasing doping. The solid and dotted lines
        are guides to the eye.}
	\protect\label{T0(x)}
\end{figure}

As seen in
Fig.~\ref{ampl(T)}, in the underdoped compounds,
at temperatures substantially higher
than $T_{c}$ superconducting fluctuations start to play an important role and
the corresponding diamagnetic signal competes the paramagnetic Van
Vleck contribution. Below $T_{0}$ the fluctuation signal
predominates the paramagnetic background and the overall response of
the sample is of diamagnetic nature.
The temperature below which
deviations from the normal paramagnetic behavior are observed, provides a
measure for the temperature regime where fluctuations are important.
However, this temperature is not well defined. On the other hand, the
temperature $T_{0}$, where the torque signal completely vanishes, is
very well defined and its value compared to $T_{c}$ yields
information about the importance of superconducting fluctuations.
Clearly, $T_{0}$ shifts closer to $T_{c}$ upon increasing the doping. The
doping dependence of $(T_{0} - T_{c})/T_{c}$ is shown in
Fig.~\ref{T0(x)}. This quantity is a good indicator for the importance
of superconducting fluctuations at different doping levels,
\noindent
since the
paramagnetic background is the same for all samples. For $x = 0.08$, at
$T_{0} \simeq 1.3 \cdot T_{c}$ the
superconducting signal already exceeds the paramagnetic
background. This means that superconducting fluctuations show up
over a wide temperature range in this doping regime. Indeed,
fluctuations are important at least up to $T \simeq 2 \cdot
T_{c}$, as shown in Fig.~\ref{ampl(T)}. On the other hand,
$T_{0}$ and $T_{c}$ can no longer be distinguished for $x = 0.18$;
as expected, fluctuations are less important in the overdoped regime, where
$\gamma$ is small.

\section{discussion}

\begin{figure}[htb]
        \centering
        \includegraphics[width=0.8\linewidth]{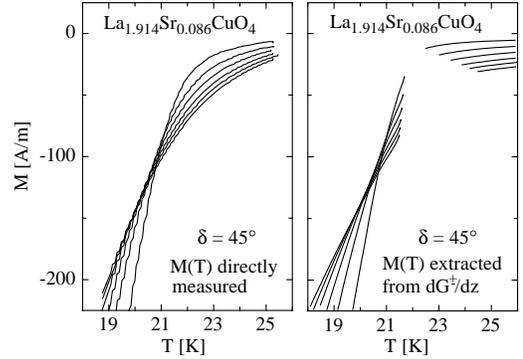}
        \caption[~]{Left panel: Magnetization $M$ as a function of
        temperature, measured on
        La$_{1.914}$Sr$_{0.086}$CuO$_{4}$ in
        different fields $B_{a} = 0.25,\ 0.50,\ 0.75,\ 1.00,\
        1.20,\mbox{ and } 1.40$ T for
        $\delta = 45^{\circ}$. Right panel: Magnetization $M$ as a function of
        temperature extracted for the same fields and the same angle from
        the experimental scaling function
        $dG^{\pm}/dz$ (see text). The agreement between the measured
        and the extracted magnetization for temperatures around
        $T^{\ast}$ is remarkable.}
	\protect\label{crossing(45deg)}
\end{figure}

Since the scaling function $dG^{\pm}/dz$ is determined experimentally, all
features showing up in
experiments are included in $dG^{\pm}/dz$. Therefore, it is
evident that the ``crossing point'' phenomenon can be deduced
from a feature contained in the scaling function. As pointed out in
Section \ref{section-crossing}, once $dG^{-}/dz$ is determined, it is
possible to predict the angular dependence of $M^{\ast}$ and $T^{\ast}$
and to  predict that the ``crossing point'' is actually a ``crossing
region'', spreading over a finite temperature interval. Moreover,
the fact that no ``crossing
point'' can be resolved for fields applied in the $ab$ plane is
explained. Note that all these predictions regarding
temperature-dependent data can be easily obtained by scaling
angular-dependent measurements, using the fact that all torque
data recorded within a not too large temperature regime around $T_{c}$
fall onto a single curve if they are scaled according to
Eq.~(\ref{tau_scaling}).
Temperature-dependent measurements performed at $\delta = 45^{\circ}$
on the La$_{1.914}$Sr$_{0.086}$CuO$_{4}$ single crystal showing the
``crossing point'' phenomenon are shown in the left panel of
Fig.~\ref{crossing(45deg)}. The right panel displays the
temperature-dependent magnetization at the same angles and fields,
extracted from the
scaling function $dG^{\pm}/dz$ which is obtained from three angular-dependent
measurements performed below $T_{c}$ (small symbols in
Fig.~\ref{dG/dz(La086)}) and from one angular-dependent
measurement performed above $T_{c}$.
\noindent
For each data point
($dG^{\pm}/dz,z$),
the corresponding temperature is calculated from $z$ using
Eq.~(\ref{z(delta)}), afterwards $M$ is calculated using
Eq.~(\ref{M}). Since
very high $z$ values are not covered by the angular-dependent
data for $T > T_{c}$, temperatures very close to $T_{c}$ are not reached.
Small deviations between the measured data and the calculated points
for temperatures very near and above $T_{c}$ are due to finite size effects
and an uncertain background correction. On the other hand, we observe a
remarkable agreement between the measured and the extracted
magnetization for
temperatures around $T^{\ast}$. For $T < \tilde{T} \simeq 19.5$ K the
magnetization curves extracted from $dG^{-}/dz$ start to deviate significantly
from the directly measured curves, because we leave
the temperature regime where the scaling approach is applicable.
This temperature region is thus estimated to be $T_{c} - \tilde{T}
\approx 0.1 \cdot T_{c}$.

The investigation of the ``crossing point'' phenomenon is one
example for the powerful possibilities of the scaling approach. Another
example regarding the melting line is given in
Ref.~\onlinecite{HoferPRB99}. Moreover, by integrating $dG^{\pm}/dz$
with respect to $z$ one gets the scaling
function of the singular part of the free energy density, Eq~(\ref{f}),
where the integration constant is determined by the condition $G(0) =
1$ (Ref.~\onlinecite{SchneiderEPJ98}). Other quantities
of interest as the specific heat can then be calculated using
thermodynamic relations.

The determination of the scaling function $dG^{\pm}/dz$ from angular-dependent
measurements is a very reliable technique, since a rotation of the sample at
fixed magnetic field and temperature allows to scan over a large $z$
regime,
$z_{\mbox{\footnotesize{min}}} < z < \gamma \cdot
z_{\mbox{\footnotesize{min}}}$,
by only changing the well controllable parameter
$\delta$ [Eq.~(\ref{z(delta)})]. One should mention, that in order to
test the expression for the
3D $XY$ free energy density [Eq.~(\ref{f})], it is much more
convenient to investigate the magnetization or the magnetic torque
than to investigate the specific heat.
This is due to the fact that taking the derivative of $f$ with respect
to the magnetic field $B_{a}$ or the angle $\delta$ (leading to the
magnetization or the magnetic torque) only results in one term,
whereas the double derivative of $f$ with respect to the temperature
leads to six terms. On the other hand, specific heat measurements
performed in zero field around $T_{c}$ yield important information about
finite size effects. \cite{SchneiderUNP}

The occurrence of a ``crossing point'' in the cuprates is ascribed to a
quasi 2D behavior of these materials.
\cite{SchneiderPC99,BulaevskiiPRL92,TesanovicPRL92}
As shown in this work, the same crossover in the behavior of the
magnetization  (i.e. a decrease of $|M|$ with increasing field for low
temperatures and an increase of $|M|$ with increasing field for high
temperatures) also exists in the 3D $XY$ model. This is seen by
recalling again the two limits $z \to 0^{-}$ and $z \to \infty$
where $M \propto \ln([\xi_{a,0}^{-}]^{2} B_{a} / \Phi_{0})$ and
$M \propto \sqrt{B_{a}}$, respectively. The square root dependence of
$dG^{\pm}/dz$ for $T \to T_{c}$ is a
characteristic property of the 3D $XY$ model. As seen from
Eq.~(\ref{tau_scaling}), a finite torque at $T_{c}$ is only possible
for $dG^{\pm}/dz \propto \sqrt{z} \propto \xi_{a,0}^{\pm} |t|^{-2/3}$.
The resulting very unusual angular dependence
of the magnetic torque near $T_{c}$ is indeed observed
[cf. Fig.~\ref{limits}(c)] and gives very
strong evidence for 3D critical fluctuations close to $T_{c}$. On the other
hand,
the low $z$
limit is mainly covered by torque data recorded for angles close to
the $ab$ plane, where the cuprates show 3D
behavior.\cite{HoferPRB99,SchneiderPC99,AegerterPRB98} However, it is not clear
whether the ``crossing region'' $T \simeq T^{\ast}$ ($z \approx 0.7$)
can be described as an exact 3D behavior, as well.
If we attribute the
``crossing point'' to a quasi 2D behavior, this means that
for intermediate $z$ values $dG^{-}/dz$ includes data recorded in a
region where quasi 2D effects become important. One would then still expect
one single 3D scaling
function $dG^{-}/dz$ for all samples at temperatures very close to $T_{c}$,
i.e. for large values of $z$. However, as already mentioned above, in this
temperature regime different $T_{c}$ distributions in the samples
lead to discrepancies of the scaling functions. In addition, the
overlap of the scaled curves occurs in the intermediate $z$ regime,
i.e. the scaling procedure is optimized for a $z$ region where the
cuprates may not show exact 3D behavior. This might explain the
systematic shift of $dG^{-}/dz$ with increasing anisotropy observed in
Fig.~\ref{dG/dz(all)}. On the other hand, deviations
from a 3D behavior are not too large since it is still possible to
perform a scaling procedure based on the 3D $XY$ model for each
sample. Note, that due to restrictions of our magnet all measurements
are performed in relatively low fields $B_{a} < 1.5$ T. For all samples
used in this work, the crossover field $B_{cr}$, above which the pancake
vortices
loose their correlation along the $c$ axis and thus show quasi 2D
behavior,\cite{SchneiderPC99,AegerterPRB98} is
larger than $2$ T. It is well possible, that for higher fields
$B_{a} > 5$ T 2D effects are much more important, and
the 3D scaling approach, restricted to a narrow temperature
regime around $T_{c}$, breaks down for $T \approx T^{\ast}$.

From Fig.~\ref{T0(x)} it is evident that fluctuation effects play a
more important role in the underdoped regime of high-$T_{c}$
systems whereas they are less important in the overdoped regime.
If we take into account that the anisotropy $\gamma$ increases in
the underdoped regime (see Table~\ref{ampl} and
Refs.~\onlinecite{ChienPC94,HoferPC98}),
that strongly underdoped compounds show 2D
scaling behavior\cite{SchneiderAPPA97} and that fluctuations are
strong in quasi 2D systems,\cite{GoldenfeldBOOK92}
our measurements are consistent with the picture that highly overdoped
high-$T_{c}$ cuprates may be treated within a 3D mean field theory,
whereas strongly underdoped compounds are dominated by (2D)
fluctuations.\cite{SchneiderAPPA97} Although an enhanced anisotropy
increases the fluctuation regime, it may shrink the temperature
regime where 3D scaling is applicable, due to enhanced
corrections to scaling.
It is worth to point out, that the occurrence of superconducting
fluctuations above $T_{c}$ clearly implies that pairs exist above $T_{c}$.
Thus, in the underdoped regime pairs are present at least up to $T
\simeq 2 \cdot T_{c}$.

In summary we applied a 3D $XY$ scaling approach to a variety of
high-$T_{c}$ cuprates with different anisotropy $\gamma$. For each
sample, scaled angular-dependent torque data recorded near $T_{c}$
fall onto the scaling function $dG^{\pm}/dz$. Starting from this scaling
function, it is possible to predict the behavior of different physical
quantities, such as magnetization, specific heat etc. close to $T_{c}$,
using the fact that all
measurements fall onto the same curve $dG^{\pm}/dz$, provided they are scaled
properly. We are able to predict
an angular-dependent ``crossing temperature'' $T^{\ast}(\delta)$ where the
magnetization $M = M^{\ast}$ seems to be independent of the applied
field. Investigations of the angular-dependent torque data recorded
at $T > T_{c}$ on
La$_{2-\mbox{\footnotesize{x}}}$Sr$_{\mbox{\footnotesize{x}}}$CuO$_{4}$
single crystals with different doping $x$ show that fluctuation
effects are more important in the underdoped regime. This is
understood in terms of an enhanced anisotropy in the underdoped
samples and supports the picture that a 3D mean field treatment of
high-$T_{c}$ materials is applicable in the highly overdoped regime,
whereas strongly underdoped compounds are dominated by (2D)
fluctuations. The
scaling functions obtained for samples with higher $\gamma$ deviate
systematically with increasing anisotropy from that obtained for the
sample with the lowest $\gamma$ value.
This may be understood by the fact
that the layered high-$T_{c}$ cuprates show a 3D-2D crossover with
decreasing temperature. This
crossover enhances the corrections to scaling and reduces the 3D
$XY$ critical regime.

\section*{acknowledgments}

The authors would like to thank K.~Kwok for providing the
YBa$_{2}$Cu$_{3}$O$_{7}$
sample. Fruitful discussions with C.~Rossel are deeply acknowledged.
This work was partly supported by the Swiss National
Science Foundation, and by NEDO and
CREST/JST (Japan). One of the authors (T.S.) would like to thank JSPS for
financial support.

\end{document}